\documentclass[aps,twocolumn,showpacs,superscriptaddress,preprintnumbers,nofootinbib,
amsmath,amssymb]{revtex4-1}

\usepackage{graphicx} 
\usepackage{tikz}
\usetikzlibrary{arrows}
\usepackage{multirow}
\usepackage[colorlinks=true,urlcolor=blue,linkcolor=blue,citecolor=blue]{hyperref}
\usepackage[normalem]{ulem}
\usepackage[capitalize]{cleveref}
\usepackage{amsfonts,amssymb}
\usepackage{comment}
\usepackage{flushend}
\usepackage[T1]{fontenc}
\usepackage{mathtools}
\usepackage{xcolor}
\usepackage{graphicx}
\usepackage{dcolumn}
\usepackage{bm}
\usepackage{flushend}

\begin{document}
\preprint{TU-1305, KEK-QUP-2026-0014, KEK-TH-2870}

\title{
High-sensitivity Ultralight Dark Matter detector: \\
Parametrically Amplified Casimir Devices }

 \begin{abstract}
We propose the use of sphere-and-plate Casimir force measurement setups to detect ultralight vector dark matter.
We demonstrate that sub-picometer signals can be parametrically amplified to be detectable.
We present a novel improvement that allows the continuous tuning of the natural frequency of the system, opening new research opportunities at the submicron scale.
We show that the improved setup can determine the dark matter mass with high accuracy, significantly expand the detectable mass range, and achieve the best sensitivity to date in the $8\times10^{-14}\sim1\times10^{-12}{\rm~ eV}$ mass range.  
 \end{abstract}

\author{Muping Chen}
\email{mpchen@post.kek.jp}
\affiliation{International Center for Quantum-field Measurement Systems for Studies of the Universe and Particles (QUP, WPI),
High Energy Accelerator Research Organization (KEK), Oho 1-1, Tsukuba, Ibaraki 305-0801, Japan}

\author{Hideo Iizuka}
\affiliation{International Center for Quantum-field Measurement Systems for Studies of the Universe and Particles (QUP, WPI),
High Energy Accelerator Research Organization (KEK), Oho 1-1, Tsukuba, Ibaraki 305-0801, Japan}

\author{Kyohei Mukaida}
\email{kyohei.mukaida@kek.jp}
\affiliation{Theory Center, IPNS, KEK, 1-1 Oho, Tsukuba, Ibaraki 305-0801, Japan}

\author{Kazunori Nakayama}
\email{kazunori.nakayama.d3@tohoku.ac.jp}
\affiliation{Department of Physics, Tohoku University, Sendai, Miyagi 980-8578, Japan
}
\affiliation{International Center for Quantum-field Measurement Systems for Studies of the Universe and Particles (QUP, WPI),
High Energy Accelerator Research Organization (KEK), Oho 1-1, Tsukuba, Ibaraki 305-0801, Japan}

\author{Burkhant Suerfu}
\email{suerfu@alumni.princeton.edu}
\affiliation{School of Physics, The University of Melbourne, Parkville, Melbourne, 3010, VIC, Australia}
\affiliation{ARC Centre of Excellence for Dark Matter Particle Physics, Australia}
\affiliation{International Center for Quantum-field Measurement Systems for Studies of the Universe and Particles (QUP, WPI),
High Energy Accelerator Research Organization (KEK), Oho 1-1, Tsukuba, Ibaraki 305-0801, Japan}

\maketitle

{\it Introduction--}
Dark matter(DM), which constitutes about 85\% of all matter in the Universe~\cite{Hinshaw_2013,Planck:2018vyg}, has so far only been observed through its gravitational interactions, and its nature remains elusive. 
Many experiments have been proposed to examine a broad mass range of DM candidates with different interactions.

As one of the promising candidates, ultralight dark matter (ULDM), with mass $\ll 1 ~{\rm eV}$, has a macroscopic coherent length and can be treated as an oscillating background field. 
It therefore exerts an oscillating force on the object carrying the corresponding dark charge~\cite{PhysRevLett.121.061102,Manley:2020mjq}. 
As a specific example, we consider a $U(1)$ gauge vector boson that couples to the difference between the baryon number and the lepton number $B-L$. This $U(1)$ gauge vector boson can be produced through mechanisms such as the misalignment mechanism~\cite{Nelson:2011sf,Nakayama:2019rhg,Nakayama:2020rka,Kitajima:2023fun} or the cosmic string network decay~\cite{PhysRevD.99.063529,Kitajima:2022lre}.
Within its coherent time, the $B-L$ force should generate a detectable, nearly monochromatic signal in precision experiments such as the equivalence principle test from the E\"{o}t-Wash group~\cite{PhysRevLett.100.041101,Wagner_2012,PhysRevD.50.3614,PhysRevD.105.042007} and the gravitational wave detectors~\cite{Pierce:2018xmy,PhysRevD.105.063030,LIGOScientific:2025ttj}.

As of today, such a signal has not been confirmed in any experiment, and new high-sensitivity experiments are in demand. One promising method for detecting ULDM is to utilize the Casimir force measurement devices, which have already been used in constraining the Yukawa force~\cite{PhysRevLett.94.240401,universe7030047,Klimchitskaya:2025ghy,PhysRevLett.116.221102}.

Casimir force measurement devices are naturally sensitive to small mechanical oscillations, due to the strong distance dependence of the Casimir force.
Displacements induced by the weak $B-L$ forces that oscillate at $\omega_0$, the natural frequency of the system, can be significantly amplified if energy is pumped into the system through the Casimir force at $2\omega_0$. This effect is called parametric amplification~\cite{PhysRevLett.67.699,Szorkovszky:2011uoa}, capable of enhancing the signal up to $10^4$-fold~\cite{Javor2021Casimir}.
Additionally, over many years of development, uncertainties in Casimir force measurements have been studied extensively~\cite{PhysRevLett.81.4549,PhysRevA.100.022508}, and built setups are available in many labs~\cite{PhysRevA.82.062512,PhysRevLett.103.040402,PhysRevLett.116.221102,10.1021/acs.nanolett.5c01101}. 

In spite of the aforementioned advantages, one significant challenge in detecting ULDM with currently available Casimir force experimental setups is that 
parametric amplification can only be achieved in a narrow band near the probe's natural frequency, thereby limiting the detection capability of each individual setup.

In this letter, we first propose the use of a sphere-and-plate Casimir force measurement device to detect the ULDM. We present a novel improvement to the currently available setups,  
which allows for the continuous tuning of the natural frequency of each setup over a considerable frequency range,
enabling the scanning over a wider detectable frequency range using a few pre-manufactured setups.
We present results for a few setups using micron-level-thickness cantilevers, showing
that best-to-date sensitivity to the $B-L$ force can be achieved in the $120 \sim 1520{\rm ~Hz}$ frequency range. 

\begin{figure}
    \centering
    \includegraphics[width=1.0\linewidth]{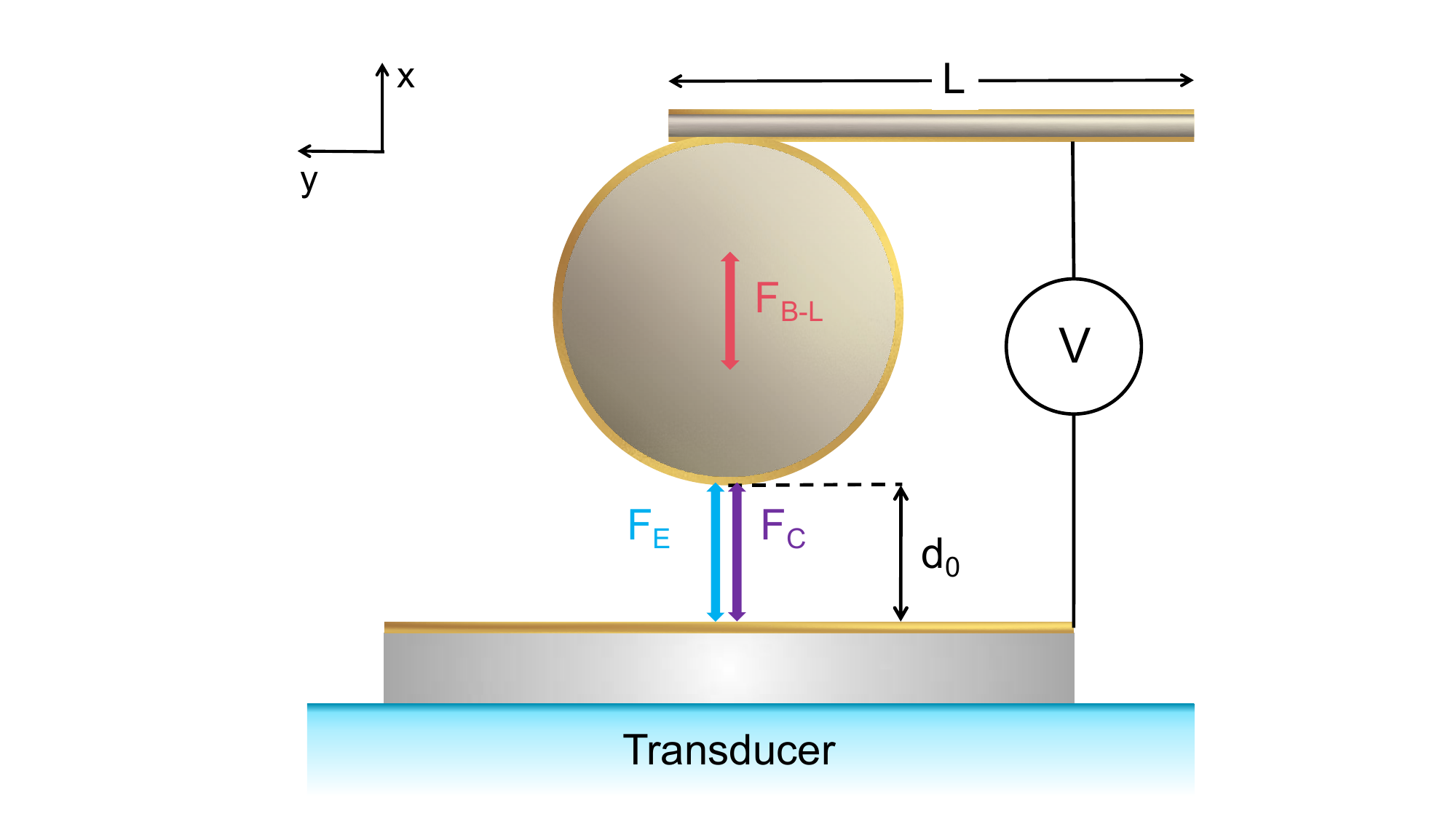}
    \caption{Schematic diagram of the Casimir ULDM detector. Both the sphere and plate are made of silicon with a $100{\rm ~nm}$ gold coating. The two surfaces are electrically connected to an external voltage source. The plate is mounted to a piezoelectric transducer, and the sphere is attached to an Au-coated ${\rm Si_3N_4}$ cantilever.}
    \label{fig:CasExp}
\end{figure}

{\it The Casimir ULDM detector--}
The basic Casimir force measuring setup (Shown in Fig.~\ref{fig:CasExp}), similar to those described in Refs.~\cite{PhysRevA.82.062512,PhysRevA.100.022508}, consists of a silicon sphere of radius $R_s$ (mm size) coated with a $100$-nm Au film, which is attached to a  Au-coated (for surface conductivity) ${\rm Si_3N_4}$ cantilever, and a plate made of a Si substrate also coated with a $100$-nm Au film, but attached to a piezoelectric transducer that can generate a controlled oscillation. The two Au surfaces are connected through an external voltage source for simultaneous calibration and parameter measurement \cite{PhysRevA.82.062512}. The whole setup is placed in a vacuum and cooled to $\sim 1 {\rm~ K}$ using liquid Helium.

At the position of the sphere, along the direction normal to the plate, which we choose to be the $x$-direction, this sphere-and-plate system is roughly equivalent to a mass-spring system with the mass of the sphere $m_s$ and the spring constant of the cantilever $k$, obeying the forced damped oscillator equation
\begin{equation}
    \label{eq:FHO}
    \ddot{x}_s+\frac{\omega_0}{Q}\dot{x}_s+\omega^2_0x_s=\frac{F_{\rm ext}}{m_s}~.
    \end{equation}
Here,  $\omega_0\equiv\sqrt{k/m_s}$ is the natural frequency and $Q$ is the quality factor of the spring, which can reach $\sim10^5$ in vacuum.

In this work, we are interested in detecting ULDMs with masses $m\leq 10^{-10}$~eV that couple to the $B-L$ charge of the sphere-cantilever oscillator probe. At the location of the earth, DM particles in the dark halo have an average velocity $v\sim 10^{-3}$ and density of $\rho_{\rm DM}=0.4~{\rm GeV/cm^3}$. With de Broglie wavelength $\lambda_{\rm DM}\sim (mv)^{-1}\simeq 2.0\times10^6~{\rm m}~[10^{-10} {\rm ~eV}/m]$, ULDMs of interest behave as classical waves and can be detected as monochromatic waves within coherent time $\tau_{\rm DM}\sim (mv^2)^{-1} = Q_{\rm DM}/m \simeq 6.6{\rm ~s}~[10^{-10} {\rm ~eV}/m]$ with $Q_{\rm DM}\sim 10^6$.
Those waves exert an external force 
\begin{equation}
    \label{eq:FB-L}
        {F}_{B-L}=\frac{1}{\sqrt{3}}gq_{B-L}E_0\cos(mt+\phi(t))~
\end{equation}
on the probe (spatially averaged) with frequency $m$ and unknown phase $\phi(t)$ (which can be treated as a constant within $\tau_{\rm DM}$),
whose amplitude depends on the unknown coupling constant $g$, the DM field energy density $E_0=\sqrt{2\rho_{\rm DM}}$, and the number of $B-L$ charge of the probe, $q_{B-L}$. 
$F_{\rm ext}=F_{B-L}$ alone causes the sphere to oscillate at frequency $m$. At resonance $m=\omega_{\rm res}=\omega_0\sqrt{1-1/4Q^2}$, we expect to observe an amplified oscillation with amplitude $|\Delta x_s|\simeq F_{B-L}Q/k$. 

The oscillation can be further amplified by parametric amplification~\cite{PhysRevLett.67.699,10.1063/1.4896732,Javor2021Casimir}. Here, we consider the parametric amplification due to the Casimir force $F_C$, which becomes significant when the separation $d<1 ~{\rm \mu m}$.
The oscillation of the sphere changes the Casimir force. For a small oscillation $\Delta x_s$, $F_C\approx F_C(x_{\rm eq})+F_C'|_{x_{\rm eq}}(\Delta x_s)$, where $x_{\rm  eq}$ is the equilibrium position of the sphere. 

To obtain parametric amplification, we shake the plate with amplitude $\Delta d$ and frequency $2\omega_0$. This shifts $x_{\rm eq}$ sinusoidally so that $x_{\rm eq}\approx x_0+\Delta d\sin(2\omega_0 t)$ and
\begin{eqnarray}
    F_C&\approx& F_C(x_0)+F_C'|_{x_0}\Delta d \sin(2\omega_0 t)\\
    &&+\left[F'_C|_{x_0}+F''_C|_{x_0}\Delta d \sin(2\omega_0 t)\right] \Delta x_s~.\notag
\end{eqnarray}
The oscillation of the sphere at $\omega_0$ can then be described by 
\begin{equation}
    \ddot{x}_s+\frac{\omega_0}{Q}\dot{x}_s+\frac{k'+\Delta k\sin(2\omega_0 t)}{m_s}x_s=g \frac{q_{B-L}E_0}{\sqrt{3}m_s}\cos(\omega_0 t+\phi),
\end{equation}
where $k'=k_0-F_C'|_{x_0}$ is the shifted spring constant, and we define $\Delta k=F_C''|_{x_0}\Delta d$.
Energy is continuously pumped into the system through $\Delta k \sin(2\omega_0 t)$, amplifying the oscillation amplitude due to force $F$ to $|\Delta x_s|=FQG/k'$ after reaching steady-state, where the phase-dependent amplification factor $G$ is (see the derivation in App.~\ref{app:Gfactor})
\begin{eqnarray}
    G(\phi)=\sqrt{\left( \frac{\cos\phi}{1+Q\Delta k/2k'}\right)^2+\left( \frac{\sin\phi}{1-Q\Delta k/2k'}\right)^2}~.
\end{eqnarray}

Despite being limited by noise and measurement precision, in principle the $[1+Q\Delta k/(2k')]^{-1}$ factor can grow to a significantly large factor as we tune $Q\Delta k/(2k')\rightarrow 1$ ($\Delta k$ is negative for $F_C$), capable of bringing a sub-picometer signal to detection. 

Effectively, by using parametric amplification, we ensure that $Q_{\rm eff}=GQ>Q_{\rm DM}$ can be achieved within $\tau_{\rm DM}$, which is hard to achieve otherwise for mechanical resonators.

To prevent Casimir pull-in (oscillation of sphere growth indefinitely and finally touches the plate, see Ref.~\cite{10.1063/1.4896732}), we require that
$k'+\Delta k\sin{(2\omega_0 t)}>0$ at all times. The minimum separation $d_{\rm min}$ and maximum oscillation $\Delta d_{\rm max}$ that prevents indefinite growth can be solved from
\begin{eqnarray}\label{eq:caspullin}
\begin{cases}
   k_0-F'_C(d_{\rm min}-\Delta d_{\rm max})=0~, \\
   k_0d_{\rm min}-F_C(d_{\rm min}-\Delta d_{\rm max})=0~. 
\end{cases} 
\end{eqnarray}
In this case, $\Delta d_{\rm max}\simeq d_{\rm min}/3.6$, and $d_{\rm min}$ depends on the cantilever used. For a commerically available $500{\rm ~\mu m}\times100{\rm ~\mu m}\times1 {\rm ~\mu m}$ cantilever,  $d_{\rm min}$ is estimated to be $\sim 110 {\rm ~nm}$.

Oscillation of the sphere is measured by an atomic force microscope (AFM) detection head with picometer sensitivity, which determines the sphere's position by detecting a laser beam reflected from the free end of the cantilever using a position-sensitive photodetector.
The voltage signal $V_{\omega_0}$ corresponding to the oscillation at  $\omega_0$ is read by a lock-in amplifier. $V_{\omega_0}$ can be translated to displacements using $V=\gamma \Delta  x$, where $\gamma$ is the optical lever sensitivity (V/m). 

The existence of $F_{B-L}$ at $m$ near $\omega_0$ leads to a phase-dependent Lorentzian signal with peak spectral density
\begin{eqnarray}
\label{eq:Sm}
    S_{FF}^{\rm DM}&=&\frac{1}{3}(g q_{B-L}E_0)^2\tau_{\rm DM}~,
\end{eqnarray}
corresponding to $|\Delta x_s|=\sqrt{S_{FF}^{\rm DM}\tau_{\rm DM}/4m_s^2\omega_0^2}$  for $Q_{\rm eff}>Q_{\rm DM}$ . 
We require $|\Delta x_s|> 1~{\rm pm}$ for a detectable signal, from which we obtain the detection threshold $S_{FF}^{\rm det}=(1{\rm ~pm})^2\tau_{\rm DM}k^2/Q_{\rm DM}^2$.
The maximum displacement due to $F_{B-L}$ is also capped from above to prevent Casimir pull-in, $|\Delta x_s|_{\rm max}<\Delta d_{\rm max}$. 

\begin{figure}
    \centering
    \includegraphics[width=0.9\linewidth]{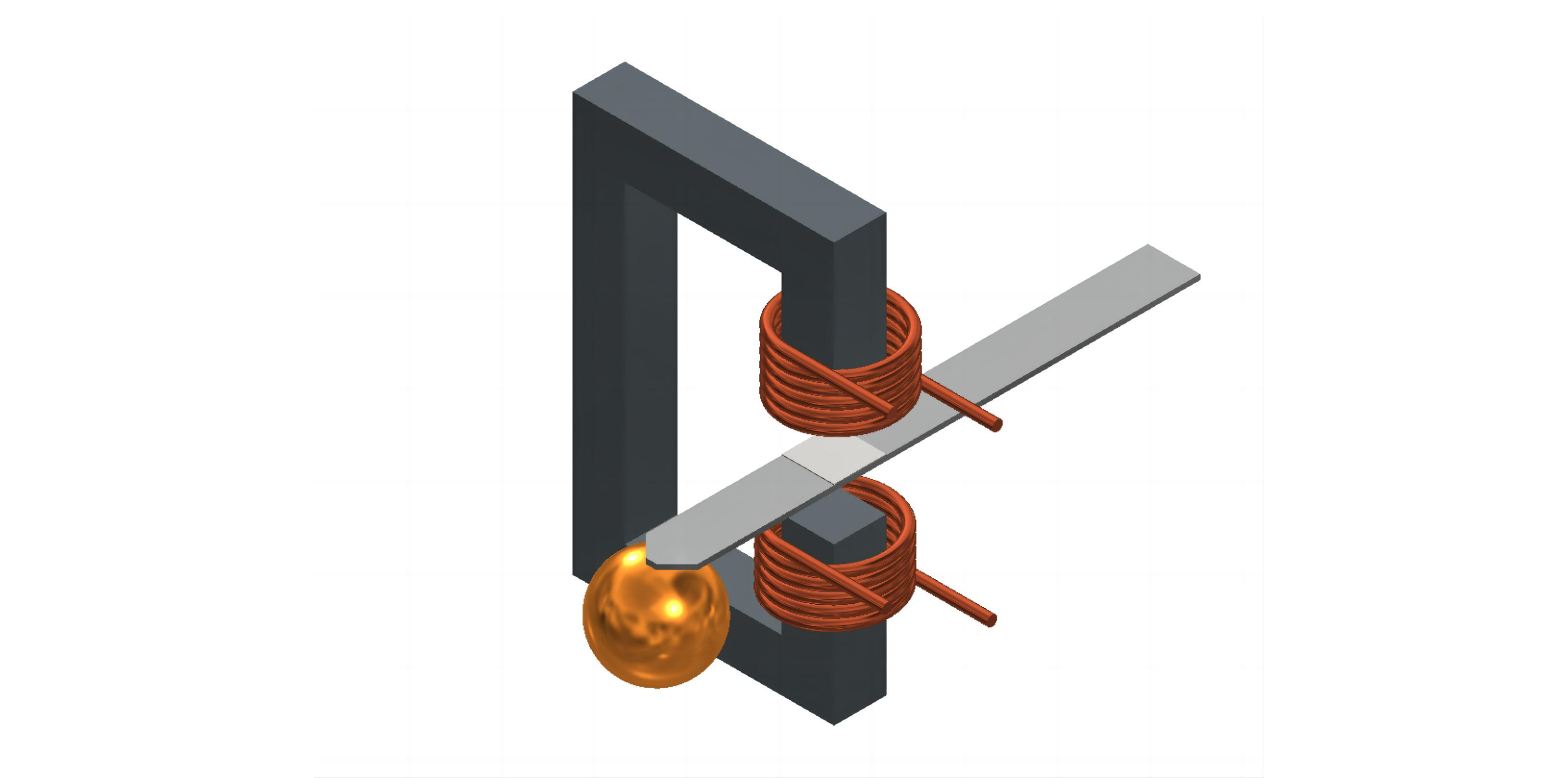}
    \caption{Drawing of the natural frequency tuning attachment. Copper wire is wound around a C-shaped metal to create a confined, adjustable magnetic field. A ferromagnetic patch (light gray) is attached to the cantilever at a chosen location, with magnetization aligned with the generated field for a controllable restoring torque.}
    \label{fig:MB-attach}
\end{figure}

{\it Tuning the resonant frequency--}
While offering great signal amplification, parametric amplification is limited to the natural frequency of the system. For the background $B-L$ force, a natural frequency tuning method is necessary to perform parameter scan over $m$. Here we first propose a novel improvement to the sphere-cantilever probe, capable of continuously adjusting the resonant frequency with negligible impact on the calibrations and measurements.

Fig.~\ref{fig:MB-attach} shows a conceptual diagram of the improved setup. 
A ferromagnetic patch (lighter gray) with magnetization $\vec{\mu}$ is placed at a chosen location in a uniform magnetic field $\vec{B}$ generated by the current-carrying coils. The magnetic field source is specifically designed to confine the magnetic field to the size of the patch, minimizing its effect on the Casimir force and the $B-L$ force measurements. The direction of $\vec{\mu}$ is set to align with $\vec{B}$ initially; misalignment between $\vec{\mu}$ and $\vec{B}$ caused by bending of the cantilever generates a restoring torque, modifying the effective spring constant, $k_{\rm eff}$. The natural frequency can be solved in terms of $m_s$, $\mu$, and $B$. Details are given in App.~\ref{app:natf}.
The magnetic field strength $B$ can be adjusted by changing the current that passes through the coils. For the purpose of increasing the ``stiffness'' of the cantilever, $B$ is not restricted from above. 

Cantilevers with different thicknesses $h$ can scan different $m$ ranges with distinct sensitivity.
The maximum mass each cantilever can hold, $m_{s,{\rm max}}$, is limited by the maximum tensile strength of the cantilever, which can reach $\sim{\rm GPa}$ at cryogenic temperatures. We perform a force simulation for ${\rm Si_3N_4}$ cantilevers using {\bf Autodesk Inventor Professional 2023} to calculate the maximum force that can be exerted on the tip of the cantilever before fracture.

The optimal detectable range for each cantilever can be calculated using $\sqrt{k_{\rm eff}/m_{s,{\rm max}}}$. The lowest detectable $m$ is obtained when $B=0$, where $k_{\rm eff}=k_0$ and $m_{\rm min}\simeq\sqrt{k_0/m_{s,{\rm max}}}$.  

The minimum $\Delta d$ required to amplify the signal above the detection threshold, $\Delta d_{\rm min}$, is also different for each setup with a different combination of sphere size and cantilever thickness, which can be solved from $QF_C''|_{x_0}\Delta d /(2k_0)=1$. $\Delta d_{\rm min}$ shall be in the tunable range $(\mathcal{O}(10^{-2} ~{\rm nm}))$ and do not cause Casimir pull-in, $\Delta d_{\rm min}<\Delta d_{\rm max}$ (see Eq.~\ref{eq:caspullin}).
Here we studied 11 cantilevers with the same area ($500 ~\mu m\times 100~ \mu m$) but different thicknesses ($1 \sim 20 ~\mu m$).
Results are listed in Tab. \ref{tab:11cant} for each cantilever.

{\it Sensitivity--}
The sensitivity is limited by the resolving power of the photo-detector, uncertainties in the experiment, and the integration time~\cite{PhysRevLett.126.061301}. 
Here, we present a sensitivity projection considering the uncertainties in measurements and three noise sources, the thermal motion of the cantilever, the shot noise in the photodetector, and the noise due to photons hitting the cantilever.
Under parametric amplification, the thermal fluctuation is also amplified, thereby being dominant, creating a signal 
\begin{eqnarray}
    S_{FF}^{\rm th}=\frac{(2k_BT)m_sm}{Q}~.
\end{eqnarray}

The uncertainties affect the sensitivity in two ways. 
First, the uncertainties restrict the maximum value of the amplification factor $G$.
As we tuning $Q\Delta k/(2k')\rightarrow 1$, in principle any sub-picometer signal can be amplified above the detection threshold. However, the uncertainties in tuning $\Delta k/k'$ limits how close $Q\Delta k/(2k')$ can approach 1, setting an upper limit on $G$. 
Second, the uncertainties determine the minimum detectable signal, $S_{FF}^{\rm n}={\rm Max}[S_{FF}^{\rm th},S_{FF}^{\rm det}]$. 

\begin{figure}
    \centering
    \includegraphics[width=0.9\linewidth]{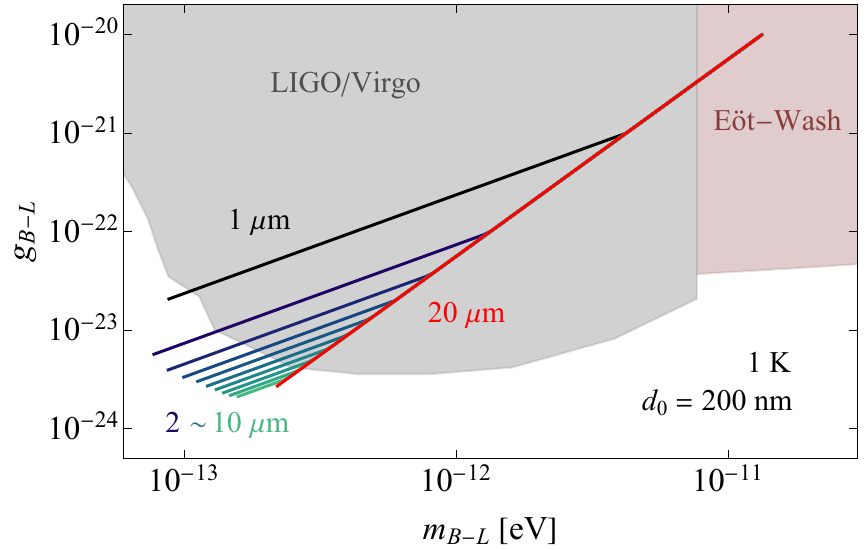}
    \caption{Sensitivity projection for 11 cantilevers for the initial separation $d_0=200 {\rm~nm}$ and the quality factor $Q=10^5$. The black line indicates the sensitivity of the commercially available $1~{\rm \mu m}$-thick cantilever. The red line indicates the sensitivity of the cantilever with 20-${\rm\mu m}$ thickness. Sensitivities for cantilevers with thickness from $2~{\rm \mu m}$ to $10~{\rm \mu m}$ are shown in colors from blue to green.
    The gray area indicates the constraint from LIGO/Virgo~\cite{Pierce:2018xmy,PhysRevD.105.063030,LIGOScientific:2025ttj} and the dark red area indicates the constraint from E\"{o}t-wash~\cite{PhysRevLett.100.041101,Wagner_2012,PhysRevD.50.3614,PhysRevD.105.042007}.}
    \label{fig:sensall}
\end{figure}

Fig.~\ref{fig:sensall} shows the estimated sensitivity assuming a unity signal-to-noise ratio, $S_{FF}^{\rm n}=S_{FF}^{\rm DM}$, measured within the coherent time $\tau_{\rm DM}$.
Sensitivities for the commercially available $1~{\rm \mu m}$-thick cantilever and the $20 ~{\rm \mu m}$-thick cantilever are shown in black and red, respectively.
Sensitivities for cantilevers with thickness from $2\sim10~{\rm \mu m}$ are shown in a color gradient from dark blue to green. All lines are shown between $\omega_{\rm min}$ (setup-dependent) and $1.34\times10^{-11}~{\rm eV}$, where the DM signal becomes undetectable.
The shaded gray region indicates the current best ULDM constraint from LIGO/Virgo~\cite{Pierce:2018xmy,PhysRevD.105.063030,LIGOScientific:2025ttj}, and the shaded dark red region indicates the constraint from E\"{o}t-wash~\cite{PhysRevLett.100.041101,Wagner_2012,PhysRevD.50.3614,PhysRevD.105.042007}.

The sensitivity improves as $m$ decreases in all setups.  The projection lines scale as $g\propto m$ in the region where $S_{FF}^{n}=S_{FF}^{\rm th}$ (detectable thermal noise), or $g\propto m^{2}$ in the region where $S_{FF}^{n}=S_{FF}^{\rm det}$ (undetectable thermal noise). The $1~{\rm \mu m}$-thick cantilever (black) has the worst sensitivity, almost entirely covered by the LIGO/VIRGO limit.
From $2~{\rm \mu m}$ to $10~{\rm \mu m}$ (blue to green), increasing the thickness of the cantilever improves sensitivity at the cost of shrinking the detectable frequency range, and all setups can probe below the LIGO/VIRGO limit between $\omega_{\rm min}$ and $\sim2\omega_{\rm min}$.
In the whole detectable range of the $20{\rm ~\mu m}$ cantilever (red),  the thermal noise is undetectable. Setups using thicker cantilevers and larger spheres will no longer increase sensitivity.

{\it Optimization--}
In addition to improving sensitivity, we also need to optimize our Casimir ULDM detector to address technical challenges we may face during data-taking and experiment-building.
 
\begin{table}[h]
    \centering
    \begin{tabular}{c|l l l l}
    \hline\hline
    $h({\rm \mu m}$)
     & $k({\rm N/m})$ & $R_{c,{\rm max}}({\rm m m})$ &$\omega_{\rm min}({\rm Hz})$ &$ g_{\rm min} {\rm @1~ K}(10^{-24})$ \\
     \hline
    2 & \quad 0.48 &\quad  1.52 &\quad  118.05&\quad 5.66\\
    4 &\quad  3.84 & \quad 2.57 &\quad  151.93&\quad 3.32\\
    9 &\quad  43.74 &\quad 4.45 &\quad  225.77&\quad 2.17\\
    20 & \quad 480.00 &\quad  7.59 &\quad  335.71&\quad 2.70\\
    \hline\hline
    \end{tabular}
    \caption{Properties for the four setups, including the spring constant $k$ of the cantilever, the silicon core radius of the heaviest sphere the cantilever can hold $R_{c,{\rm max}}$, the minimum detectable frequency $\omega_{\rm min}$, and the lowest $g$, $g_{\rm min}$, reachable with $\tau=\tau_{\rm DM}$ and $ T=1 {\rm ~K}$.} 
    \label{tab:4cant}
\end{table}

The sensitivity worsens as frequency increases. To achieve optimal sensitivity at each frequency, we should, in principle, use the thickest cantilever allowed at every frequency.
However, all setups that can probe below the LIGO/VIRGO limit are not commercially available and have to be fabricated, and the gain of switching to a thicker cantilever diminishes for thicker cantilevers (see Fig.~\ref{fig:sensall}).
To be cost-efficient, 
here we select four sphere-cantilever combinations (listed in Tab.~\ref{tab:4cant}), ordered by their minimum detectable frequency, $\omega_{\rm min}$, with two consecutive $\omega_{\rm min}$ roughly having the same ratio. The combined sensitivity projection of the four setups is obtained by overlapping their individual sensitivity projections.

The sensitivity can be further improved by lowering the temperature,  since $F_{\rm min}\propto T^{1/2}$. Ideally, we can place the setups in a dilution refrigerator (DR) and lower the temperature to $\sim 10~{\rm mK}$. This experiment would be very challenging to perform. To our knowledge, it has not been done yet. Here, we show an optimistic projection of what such an experiment could achieve.

\begin{figure}
    \centering
    \includegraphics[width=0.9\linewidth]{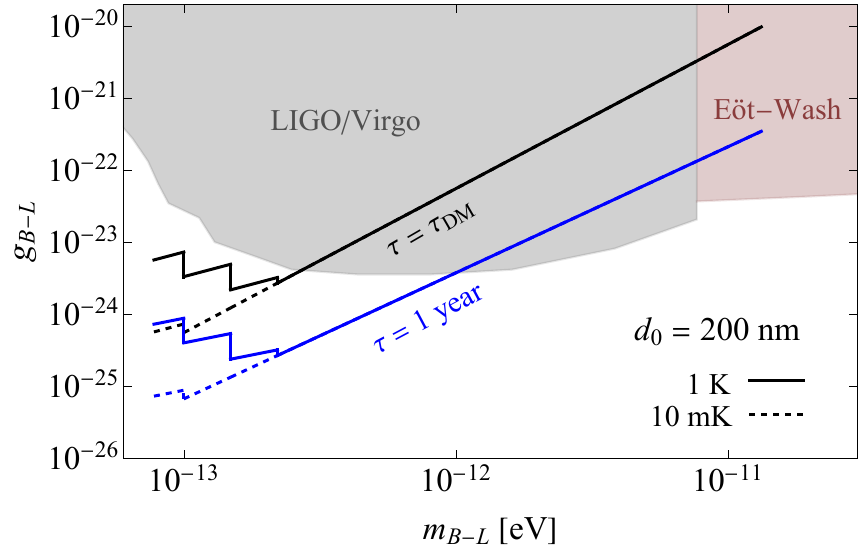}
    \caption{Combined sensitivity projection for the four setups shown in Tab.~\ref{tab:4cant}, with $d_0=200~{\rm \mu m}$ and $Q=10^5$. Shaded regions indicate the same constraints as shown in Fig.~\ref{fig:sensall}. Sensitivity for $\tau=\tau_{\rm DM}$ and $\tau=1{\rm ~yr}$ are shown in black and blue, respectively. The solid lines indicate the sensitivity reachable by cooling the system to 1 K using liquid He, and the dashed lines show the projection for further experiments done in DR at 10 mK.}
    \label{fig:senspick}
\end{figure}

The sensitivity also improves with longer integration time. Although the measurement time cannot exceed the coherent time, we can improve the sensitivity by $N^{1/4}$ through averaging $N$ periodogram measurements (each with duration $\tau_{\rm DM}$) over a total integration time $\tau=N\tau_{\rm DM}$ \cite{PhysRevLett.126.061301}. With $\tau_{\rm DM}\simeq\mathcal{O}({\rm min})\sim\mathcal{O}({\rm hr})$, 1-yr integration time can increase the sensitivity by $10\sim 30$ times. The downside of a long measurement time is that the natural frequency of each setup has to be kept the same over a long period of time, thus the advantage of having a tunable natural frequency (i.e., a wider scannable frequency range with high sensitivity) diminishes.

Fig.~\ref{fig:senspick} shows the combined sensitivity of the four setups listed in Tab.~\ref{tab:4cant}, with $Q=10^5$ and $d_0=200{~\rm nm}$, capable of scanning over the $\sim(7.8\times 10^{-14},1.3\times10^{-11}){\rm ~eV}$ mass range. At 1 K (solid lines), within the scannable mass range, the combined sensitivity is below the LIGO/VIRGO limit in $\sim(7.8\times 10^{-14},2.7\times10^{-13}){\rm ~eV}$ for $\tau=\tau_{\rm DM}$ (black), and in $\sim(7.8\times 10^{-14},1.0\times10^{-12}){\rm ~eV}$ for $\tau=1 {\rm ~yr}$ (blue). For both measurement times, we achieve the current best sensitivity in the corresponding mass ranges, reaching (the lowest $g$) $g_{\min}\sim2.17\times10^{-24}$ and $g_{\rm min}\sim2.36\times10^{-25}$, respectively. 

If the experiment can be performed at 10 mK in a DR (dashed lines), the sensitivity improves by an order of magnitude for the 2 $\mu m$-thick cantilever, but the effectiveness diminishes for larger $m$ until it completely vanishes where the thermal noise becomes undetectable. $g_{\rm min}$ reaches the lowest value of $\sim5.52\times10^{-25}$ in the measurement time $\tau_{\rm DM}$ (black) and can reach $\sim6.64\times10^{-26}$ in a measurement of a year (blue).

{\it Further discussion--}
In this letter, we have so far focused on our Casimir ULDM detector's capability to achieve high sensitivity. 
In addition to that, our Casimir ULDM detector can determine the mass of the $U(1)$ gauge vector boson with high accuracy. 
The mass can be constrained to $\omega_0\pm1/2\tau_{\rm DM}$ upon successful detection of a signal at a chosen $\omega_0$.
Also, our novel improvement not only makes implementing the Casimir ULDM detector truly feasible, but also opens new research opportunities for other high-sensitivity tests at the sub-micron length scale, such as fifth-force detection and axion searches. 

{\it Acknowledgement--}
This work was supported by World Premier International Research Center Initiative (WPI), MEXT, Japan. We acknowledge support by JSPS KAKENHI (Grant Number 24K07010 [KN], 26K00695 [KN], 26H00403 [KN]). This research was supported by the Australian Government through the Australian Research Council Centre of Excellence for Dark Matter Particle Physics~(CDM, CE200100008).
\appendix
\section{Cantilever property table}
\label{app:cant-tab}
\begin{table}[h]
    \centering
    \begin{tabular}{c|r r r r l}
    \hline\hline
    $h$(${\rm \mu m}$)
     & $k({\rm N/m})$ & $R_{c,{\rm max}}({\rm m m})$ &$\omega_{\rm min}({\rm Hz})$ &$\Delta d_{\rm min}({\rm nm})$\\
     \hline
    1 & \quad 0.06 &\quad  0.70 &\quad  134.01&\quad 0.03\\
    2 & \quad 0.48 &\quad  1.52 &\quad  118.05&\quad 0.13\\
    3 & \quad 1.62 &\quad  2.11 &\quad  133.25&\quad 0.31\\
    4 & \quad 3.84 &\quad  2.57 &\quad  151.93&\quad 0.59\\
    5 & \quad 7.50 &\quad  2.97 &\quad  170.91&\quad 1.00\\
    6 & \quad 12.96 &\quad  3.38 &\quad  185.70&\quad 1.53\\
    7 & \quad 20.58 &\quad  3.75 &\quad  200.19&\quad 2.19\\
    8 & \quad 30.72 &\quad  4.11 &\quad  212.95&\quad 2.98\\
    9 &\quad  43.74 & \quad 4.45 &\quad  225.77&\quad 3.92\\
    10 &\quad  60.00 &\quad  4.73 &\quad  240.92&\quad 5.05\\
    20 & \quad 480.00 &\quad  7.59 &\quad  335.71&\quad 25.20\\
    \hline\hline
    \end{tabular}
    \caption{Properties for the 11 sphere-cantilever combinations we studied, including the thickness $h$ of the cantilever, the spring constant $k$ of the cantilever, the Si core radius of the heaviest sphere the cantilever can hold $R_{c,{\rm max}}$, the minimum detectable frequency $\omega_{\rm min}$, and the smallest shaking amplitude needed $\Delta d_{\rm min}$.} 
    \label{tab:11cant}
\end{table}
\section{ Derivation of the amplification factor}
\label{app:Gfactor}
A parametrically amplified mechanical oscillator can be described by
\begin{eqnarray}
\label{eq:gen-param-amp}
    \ddot{x}+\frac{\omega_0}{Q}\dot{x}+\omega_0^2[1+r\sin(2\omega_0t)]x=\frac{F_0}{m}\cos(\omega_0t+\phi)~
\end{eqnarray}
where the ratio $r=\Delta k/k'$.

After the system reaches the steady-state, we expect the amplitude of the oscillation to stop changing, and thus an ansatz in the form
\begin{eqnarray}
    x(t)=x_1\cos(\omega_0t)+x_2\sin(\omega_0t).
\end{eqnarray}
Plugging into Eq.~(\ref{eq:gen-param-amp}), we obtain the cosine component ($\cos(\omega_0t)$) equation
\begin{eqnarray}
\label{eq:coseq}
    \frac{\omega_0^2}{Q}x_2+\frac{r\omega_0^2}{2}x_2=\frac{F_0}{m}\cos\phi
\end{eqnarray}
and the sine component ($\sin(\omega_0t)$) equation
\begin{eqnarray}
\label{eq:sineq}
    \frac{\omega_0^2}{Q}x_1-\frac{r\omega_0^2}{2}x_1=\frac{F_0}{m}\sin\phi~.
\end{eqnarray}
The two coefficients can be solved from Eq.~(\ref{eq:coseq}) and Eq.~(\ref{eq:coseq}), yielding
\begin{eqnarray}
&x_1&=\frac{F_0Q}{m\omega_0^2}\left[ \frac{\sin\phi}{1-(rQ/2)} \right]~;\\
&x_2&=\frac{F_0Q}{m\omega_0^2}\left[ \frac{\cos\phi}{1+(rQ/2)} \right]~.\notag
\end{eqnarray}
The total amplitude 
\begin{eqnarray}
    |\Delta x|&=&\sqrt{x_1^2+x_2^2}\\
    &&=\frac{F_0Q}{m\omega_0^2}\left[\sqrt{\left( \frac{\cos\phi}{1+rQ/2}\right)^2+\left( \frac{\sin\phi}{1-rQ/2}\right)^2}\right]\notag
\end{eqnarray}
where we identify the term in the bracket as $G(\phi)$.

\section{Natural frequency}
\label{app:natf}
In this section, we present the calculation of the fundamental frequency, or the lowest natural frequency.
The fundamental frequency of the cantilever can be obtained by solving the beam equation
\begin{eqnarray}
\label{eq:beam}
    EI\frac{\partial ^4 v}{\partial y^4}+\rho A\frac{\partial ^2v}{\partial t^2}=0~,
\end{eqnarray}
where $v$ represents the local curvature of the beam, $E$ is the elastic modulus, $I$ is the moment of inertia, $\rho$ is the density of the beam, and $A$ is the cross-sectional area of the beam.

For the original Casimir setup, the boundary conditions are $v=0$, $v'=0$ at the fixed-end ($y=0$) and $v''=0$, $EI v'''=m_s\ddot{v}$ at the free-end ($y=L$). 
The restoring torque introduces a new boundary condition at its center of mass location of the attachment, $EI(v''_- -v_+'')=-\mu B{v}'_-$, where $v_-$ and $v_+$ denote the local curvature approaching from the left and from the right.
Together with the continuity condition $v_-=v_+$, $v_-'=v_+'$, and $v_-'''=v_+'''$,
the fundamental frequency can then be solved using Eq.~(\ref{eq:beam}) in terms of the variable $\mu B$.

Separating the variable, $v(y,t)=Y(y)T(t)$, Eq.~(\ref{eq:beam}) can be separated into
\begin{eqnarray}
\label{eq:beam-T}
    \frac{\partial^2 T}{\partial t^2}=-\omega^2 T;\quad
    \frac{\partial^4 Y}{\partial y^4}=k^4Y~,
\end{eqnarray}
where $k^4=\omega^2\rho A/(EI)$. The solution to the time equation is the regular sinusoidal function in time. For the Y equation, the general solution takes the form
\begin{eqnarray}
    Y(y)&=&C_1\cos{(ky)}+C_2\sin{(ky)}\\&&+C_3\cosh{(ky)}+C_4\sinh{(ky)}\notag~.
\end{eqnarray}
The beam can be treated as two separate beams, separated at the attachment location, so we have a set of 8 unknown coefficients ${\bf c}$. The 8 boundary conditions given above can be written as a matrix equation in terms of these coefficients, $A{\bf c}^T={\bf 0}$, where
\begin{equation}
    {\bf c}=\{ C_1^-,C_2^-,C_3^-,C_4^-, C_1^+,C_2^+,C_3^+,C_4^+ \}~.
\end{equation}
Here, we denote the location of the attachment as $nL$ and define $z=kL$, then elements in $A$ can be written in terms of $z$, $n$, $\mu B$, and $m_s$. For a given $(m_s,\mu B,n)$, the variable $z$ can be solved from $\det(A)=0$, from which we obtain the fundamental frequency $\omega=(z/L)^2(EI/\rho A)^{1/2}$.

\bibliography{reference}
\end{document}